# Hermite-Chebyshev pseudospectral method for inhomogeneous superconducting strip problems and magnetic flux pump modeling

**Vladimir Sokolovsky[1] and Leonid Prigozhin[2]**

[1]Physics Department, Ben-Gurion University of the Negev, Beer Sheva 84105, Israel
[2]J. Blaustein Institutes for Desert Research, Ben-Gurion University of the Negev, Sde Boqer Campus 84990, Israel

E-mail: sokolovv@bgu.ac.il and leonid@bgu.ac.il

**Abstract**

Numerical simulation of superconducting devices is a powerful tool for understanding the principles of their work and improving their design. We present a new pseudospectral method for two-dimensional magnetization and transport current superconducting strip problems with an arbitrary current-voltage relation, spatially inhomogeneous strips, and strips in a nonuniform applied field. The method is based on the bivariate expansions in Chebyshev polynomials and Hermite functions. It can be used for numerical modeling magnetic flux pumps of different types and investigating AC losses in coated conductors with local defects. Using a realistic two-dimensional version of the superconducting dynamo benchmark problem as an example, we showed that our new method is a competitive alternative to finite element methods.

## 1. Introduction

Along with finite differences and finite elements, spectral methods are one of the three main numerical techniques for solution of partial differential and integral equations. These methods typically present an approximate solution as a series of some basis functions, such as trigonometric functions or orthogonal polynomials [1, 2]. If the exact problem solution is smooth, convergence of these expansions can be exponentially fast [3]. Spectral methods, in which spatial collocation is used to determine the expansion coefficients, are called pseudospectral and are especially convenient for variable-coefficient and nonlinear problems.

Recently, the Chebyshev polynomial expansions have been employed, see [4], for a very accurate and efficient solution of the superconductivity problems formulated as one-dimensional (1D) integrodifferential equations with a Cauchy-type singular kernel or as a system of such equations (infinite superconducting strips with or without transport current in a uniform field, stacks of strips, and pancake coils). A similar approach was used in [5] for fast numerical solution of the superconducting dynamo benchmark problem [6].

To simplify the dynamo problem and make it 1D, it is typically assumed (see [5, 6] and the references therein) that the superconducting strip (the dynamo stator) is long and the external magnetic field is produced by an infinitely long rotating permanent magnet. A more realistic model should account for the local character of an applied magnetic field and be two-dimensional (2D). Recently, a 2D dynamo pump problem was solved in [7] by two different numerical methods, the mixed finite element method and the fast Fourier transform (FFT) based method; another finite element method was used in [8]. Modeling the 2D loop currents, induced in a superconducting strip by a local (significant only in a finite area of the strip) nonstationary magnetic field, is needed also for simulating the traveling wave pumps [9]. In this work we derive a new pseudospectral method for such 2D problems.

The method can be used also for another actual problem: investigating the influence of a local inhomogeneity in a coated conductor strip. Effect of such inhomogeneities on the current density and loss cannot be studied in the frame of a 1D model.

While Chebyshev polynomial expansions are known to be convenient for solving 1D integral equations with the Cauchy and logarithmic type singular kernels (see, e.g., [10-12]), no similar approach has been developed for the singular kernel of the Green-function-based 2D integral formulation. To circumvent this difficulty, we apply the Fourier transform in the along-strip direction. For each wave number, the 2D integral equation is represented in the Fourier space by a 1D singular integral equation written for the transverse coordinate; we use expansions in Chebyshev polynomials to efficiently solve these equations. To employ this approach, we need an efficient numerical implementation of the direct and inverse Fourier transforms. Typically, this is realized by replacing the infinite axis by a sufficiently long finite interval and using the FFT algorithm on a uniform mesh. A different method we explore in this work is based on expansions in the Hermite functions which diagonalize the Fourier transform. An approximate solution to the 2D integrodifferential problem is, therefore, sought in our work at each moment in time as a bivariate Hermite-Chebyshev expansion. The method of lines is employed for integration in time.

By using the Hermite functions we avoided the strip cutoff error but, nevertheless, cannot expect an exponentially fast convergence: unlike the Hermite functions, which become exponentially small at infinity, our problem solution behaves only as a negative power. We found, however, that the method is accurate and more efficient than the methods in [7]. Comparing to [7], we have also extended our new method to problems with field-dependent current-voltage relations, characterizing the superconducting material, and to problems for inhomogeneous superconducting strips with a transport current.

Implementation of spectral methods often seems complicated but is usually based on a few algorithms for interpolation, differentiation, integration, and Fourier transform. For reader's convenience we describe in detail the algorithms for these linear operations on the expansions employed in our work.

All numerical simulations were performed in Matlab R2020b on a PC with the Intel(R) Core(TM) i7-10700 CPU @ 2.90GHz and 64 GB RAM.

## 2. Thin strip in a nonuniform field

Neglecting strip thickness, we present the strip of the width $2a$ as a 2D domain in the plane $z = 0$:

$$\Omega = \{(x, y) : |x| < \infty, |y| \leq a\}.$$

By the Faraday law

$$\mu_0 \dot{h}_z = -\nabla \times \boldsymbol{e}, \tag{1}$$

where $\mu_0$ is the magnetic permeability of vacuum, $h_z$ is the normal to strip component of the magnetic field $\boldsymbol{h}$, $\dot{f}$ denotes $\partial_t f$, $\boldsymbol{e} = (e_x, e_y)$ is the parallel to strip electric field component, and the 2D curl $\nabla \times \boldsymbol{e} = \partial_x e_y - \partial_y e_x$. Let the strip be characterized by a current-voltage relation

$$\boldsymbol{e} = \rho \boldsymbol{j}, \tag{2}$$

where $\boldsymbol{j}$ is the sheet current density and the nonlinear resistivity $\rho$ is defined, e.g., as

$$\rho = e_0(|\boldsymbol{j}|/j_c)^{n-1}/j_c \qquad (3)$$

with the critical sheet current density $j_c$ and power $n$; here $e_0 = 10^{-4}$ Vm$^{-1}$. Typically, $j_c$ depends on the magnetic field and can also be spatially inhomogeneous.

Let us assume first that no transport current is applied and the nonuniform external field $\boldsymbol{h}^e$ is essential in a bounded domain only, i.e., tends to zero sufficiently fast as $x \to \pm\infty$. Since $\nabla \cdot \boldsymbol{j} = 0$ in $\Omega$ and $j_y = 0$ at $y = \pm a$, we can introduce a stream (magnetization) function $g(x, y, t)$ such that $\boldsymbol{j} = \bar{\nabla} \times g$ (i.e. $j_x = \partial_y g$, $j_y = -\partial_x g$) and the boundary conditions are $g|_{y=\pm a} = 0$. By the Biot-Savart law, in the strip plane $z = 0$ we have

$$h_z = h_z^e + \nabla \times \int_\Omega G(\boldsymbol{r} - \boldsymbol{r}') \bar{\nabla}' \times g(\boldsymbol{r}', t) \, d\boldsymbol{r}', \qquad (4)$$

where $\boldsymbol{r} = (x, y)$, $h_z^e(\boldsymbol{r}, t)$ is the normal to strip component of the external magnetic field $\boldsymbol{h}^e$, and $G(\boldsymbol{r}) = (4\pi|\boldsymbol{r}|)^{-1}$ is the Green function. Using (1)-(4) we obtain

$$-\nabla \times \int_\Omega G(\boldsymbol{r} - \boldsymbol{r}') \bar{\nabla}' \times \dot{g}(\boldsymbol{r}', t) \, d\boldsymbol{r}' = \dot{h}_z^e + \mu_0^{-1} \nabla \times \boldsymbol{e},$$

$$\boldsymbol{e} = e_0(|\bar{\nabla} \times g|/j_c)^{n-1} \bar{\nabla} \times g / j_c, \qquad (5)$$

$$g|_{y=\pm a} = 0.$$

This is the evolutionary problem for the magnetization function we solve. An initial condition should be also supplied; in simulations we assumed $g|_{t=0} = 0$.

If the electric field $\boldsymbol{e}$ is significant only in a finite part of the strip, e.g., in the stator strip of a magnetic flux pump, the generated width-averaged voltage can be calculated as

$$V(t) = \frac{1}{2a} \int_\Omega e_x(\boldsymbol{r}, t) d\boldsymbol{r}. \qquad (6)$$

Below, we consider also problems with a given nonzero transport current $I(t)$. For simulations we will use dimensionless variables,

$$\hat{\boldsymbol{r}} = \frac{\boldsymbol{r}}{a}, \quad \hat{\boldsymbol{j}} = \frac{\boldsymbol{j}}{j_c^0}, \quad \hat{j}_c = \frac{j_c}{j_c^0}, \quad \hat{\boldsymbol{h}} = \frac{\boldsymbol{h}}{j_c^0}, \quad \hat{I} = \frac{I}{aj_c^0},$$

$$\hat{g} = \frac{g}{aj_c^0}, \quad \hat{\boldsymbol{e}} = \frac{\boldsymbol{e}}{e_0}, \quad \hat{t} = \frac{e_0}{a\mu_0 j_c^0} t, \quad \hat{V} = \frac{V}{ae_0},$$

where $j_c^0$ is a characteristic value of the sheet critical current density, e.g., the value at zero field; the strip domain in the dimensionless variables is $\hat{\Omega} = \{|\hat{x}| < \infty, |\hat{y}| \leq 1\}$.

Omitting the "hat" sign we rewrite (5)-(6) in the new variables:

$$-\nabla \times \int_\Omega G(\boldsymbol{r} - \boldsymbol{r}') \bar{\nabla}' \times \dot{g}(\boldsymbol{r}', t) \, d\boldsymbol{r}' = \dot{h}_z^e + \nabla \times \boldsymbol{e}, \qquad (7)$$

$$\boldsymbol{e} = (|\bar{\nabla} \times g|/j_c)^{n-1} \bar{\nabla} \times g / j_c, \qquad (8)$$

$$V = \frac{1}{2}\int_\Omega e_x(\mathbf{r},t)\mathrm{d}\mathbf{r} \tag{9}$$

with $g|_{y=\pm 1}=0$. Presenting the first term in equation (7) as a convolution with respect to $x$,

$$-\nabla\times\int_\Omega G(\mathbf{r}-\mathbf{r}')\bar{\nabla}'\times\dot{g}(\mathbf{r}',t)\mathrm{d}\mathbf{r}' =$$

$$\int_{-1}^{1}\partial_x G(x,y-y')*\partial_x \dot{g}(x,y',t)\mathrm{d}y' +$$

$$\int_{-1}^{1}\partial_y G(x,y-y')*\partial_{y'} \dot{g}(x,y',t)\mathrm{d}y',$$

we apply the Fourier transform $\tilde{f}(k,y,t) = F[f(x,y,t)] := \int_{-\infty}^{\infty} f(x,y,t)e^{-ikx}\mathrm{d}x$. This yields

$$-k^2\int_{-1}^{1}\tilde{G}(k,y-y')\tilde{\dot{g}}(k,y',t)\mathrm{d}y' +$$
$$\int_{-1}^{1}\partial_y\tilde{G}(k,y-y')\partial_{y'}\tilde{\dot{g}}(k,y',t)\mathrm{d}y' = \tilde{Z} \tag{10}$$

with $Z = \dot{h}_z^e + \partial_x e_y - \partial_y e_x$ and (see [13], 9.6.21 and 9.6.27)

$$\tilde{G}(k,y) = \frac{1}{4\pi}\int_{-\infty}^{\infty}\frac{e^{-ikx}\mathrm{d}x}{\sqrt{x^2+y^2}} = \frac{1}{2\pi}K_0(|yk|),$$

$$\partial_y\tilde{G}(k,y) = -\frac{|k|}{2\pi}K_1(|yk|)\mathrm{sign}(y),$$

where $K_j$ is the modified Bessel function of the second kind of order $j$. These integral kernels are singular: for $s\to 0$ we have $K_0(|s|)\simeq -\ln(|s|/2)$ and $K_1(|s|)\simeq 1/|s|$ ([13], 9.8.5 and 9.8.7, respectively). Introducing the functions

$$\mathcal{A}(k,y) = \tilde{G}(k,y) + \frac{1}{2\pi}\ln\frac{|ky|}{2} =$$
$$\frac{1}{2\pi}\left[K_0(|ky|) + \ln\frac{|ky|}{2}\right], \tag{11}$$

$$\mathcal{B}(k,y) = -\left[\partial_y\tilde{G}(k,y) + \frac{|k|}{2\pi}\frac{1}{|ky|}\mathrm{sign}(y)\right] =$$
$$\frac{|k|}{2\pi}K_1(|ky|)\mathrm{sign}(y) - \frac{1}{2\pi y} \tag{12}$$

we rewrite (10) singling out the kernel singularities at $y=y'$:

$$k^2 \int_{-1}^{1} \left\{ \mathcal{A}(k, y-y') - \frac{1}{2\pi} \ln\left(\frac{|k|}{2} |y-y'|\right) \right\} \tilde{g}(k, y', t) \,dy' +$$
$$\int_{-1}^{1} \left\{ \mathcal{B}(k, y-y') + \frac{1}{2\pi} \frac{1}{y-y'} \right\} \partial_{y'} \tilde{g}(k, y', t) \,dy' = -\tilde{Z}. \quad (13)$$

For problems with a transport current $I(t) \neq 0$ this formulation should be modified. It is also desirable to relax our assumption about the behavior of the external field at infinity: we now assume the nonuniform external field $\boldsymbol{h}^e(\boldsymbol{r}, t)$ tends to a uniform one, $\boldsymbol{h}^{e,\infty}(t)$, as $x \to \pm\infty$. We further assume the strip inhomogeneity, if present, is localized in a finite strip part. Hence, for the far away values of $x$, the sheet current density $\boldsymbol{j}(\boldsymbol{r}, t)$ should be close to $\boldsymbol{j}^\infty = (j_x^\infty(y, t), 0)$, the solution to the 1D problem for an infinite homogeneous strip with the same transport current $I(t)$ and the external field $\boldsymbol{h}^{e,\infty}(t)$. This 1D problem can be solved numerically by different methods; we used the Chebyshev spectral method described in Appendix C. Clearly, $\boldsymbol{j}^* = \boldsymbol{j} - \boldsymbol{j}^\infty$ satisfies the zero-divergence condition, and we can define the corresponding stream function $g^*$, equal to zero at the strip boundaries $y = \pm 1$. Equation (7) yields

$$-\nabla \times \int_\Omega G(\boldsymbol{r} - \boldsymbol{r}') \bar{\nabla}' \times \dot{g}^*(\boldsymbol{r}', t) \,d\boldsymbol{r}' = Z^*, \quad (14)$$

where $Z^* = \dot{h}_z^e - \dot{h}_z^{e,\infty} + \nabla \times (\boldsymbol{e} - \boldsymbol{e}^\infty)$, and $\boldsymbol{e}^\infty$, $\boldsymbol{e}$ are the parallel to film electric field components determined by (8) for the current densities $\bar{\nabla} \times g = \boldsymbol{j}^\infty$ and $\bar{\nabla} \times g = \boldsymbol{j}^\infty + \bar{\nabla} \times g^*$, respectively. The new variable $g^*$ and the right-hand-side $Z^*$ tend to zero at infinity and equation (13) holds for their Fourier transforms. Below, we present our numerical scheme for computing the magnetization function $g$; computing $g^*$ is similar.

### 3. Numerical method

To use the method of lines we need to discretize the derived evolutionary problem only in space and obtain a system of ordinary differential equations (ODE), which is then integrated in time using a standard ODE solver (we used the Matlab *ode15s* solver with the default setting and automatic accuracy control). For simplicity, we describe first our Hermite-Chebyshev numerical scheme for problems with a field-independent critical sheet current density $j_c$. Then, for problems with a field-dependent current-voltage relation, we derive also a Hermite-Chebyshev method for computing the current-induced magnetic field.

*3.1 Solution of integrodifferential equation* (13).

Let the function $g$ be known at time $t$. For each wave number $k$ the 1D integrodifferential equation (13) for $\tilde{g}$ can be efficiently solved using the Chebyshev polynomials of the first and second kind, $T_m(y)$ and $U_m(y)$, respectively, orthogonal on $[-1, 1]$ with the corresponding weights [10]:

$$\int_{-1}^{1} \frac{T_l(y)T_m(y)}{\sqrt{1-y^2}} dy = \begin{cases} 0 & l \neq m \\ \pi & l = m = 0 \\ \pi/2 & l = m > 0 \end{cases} \tag{15}$$

$$\int_{-1}^{1} \sqrt{1-y^2} U_l(y)U_m(y) dy = \begin{cases} 0 & l \neq m \\ \pi/2 & l = m \end{cases} \tag{16}$$

To compute the right-hand-side $\tilde{Z} = F[Z]$ and find $\dot{g} = F^{-1}\left[\tilde{\dot{g}}\right]$ we need to implement also the Fourier transform (with respect to $x$) and its inverse numerically; for this we use the interpolating expansions in Hermite functions $\Psi_j$. These functions (see [14] for a comprehensive presentation of their properties) are expressed via the Hermite polynomials $H_j$ as $\Psi_j(x) = \left(\pi^{1/4}\sqrt{2^j j!}\right)^{-1} e^{-x^2/2} H_j(x)$, form a basis in $L^2(R^1)$, satisfy the orthogonality relation $\int_{-\infty}^{\infty} \Psi_i \Psi_j dx = \delta_{ij}$ (the Kronecker delta), and can be efficiently computed using the recurrence relation

$$\Psi_0(x) = \pi^{-1/4} e^{-x^2/2}, \quad \Psi_1(x) = \pi^{-1/4}\sqrt{2} x e^{-x^2/2},$$
$$\Psi_{j+1}(x) = x\sqrt{\frac{2}{j+1}} \Psi_j(x) - \sqrt{\frac{j}{j+1}} \Psi_{j-1}(x), \quad j \geq 1. \tag{17}$$

The Fourier transform of $\Psi_j(x)$ is

$$F[\Psi_j(x)] = \sqrt{2\pi}(-i)^j \Psi_j(k), \tag{18}$$

which makes transitions from an expansion in the Hermite functions to its direct or inverse Fourier transform an elementary task. As is typical for approximations in an infinite domain, it is better to use expansions in scaled functions, $\Psi_j(x/L)$, where the scaling factor $L$ ensures the interpolation nodes are distributed in the domain where the approximated function values are significant (see [1, 15]). For problems considered in this work this domain is several times longer than the rotating permanent magnet in the dynamo pump problem, or than the local strip inhomogeneity, effect of which is simulated.

The choice of bivariate Hermite-Chebyshev interpolating expansions determines our choice of an $(N+1)\times(M+1)$ grid of collocation nodes in the domain $\Omega$. To eliminate the Runge phenomenon (bad polynomial approximation between the collocation points) and to satisfy exactly the boundary conditions, for $-1 \leq y \leq 1$ we use the Chebyshev points of the second kind, $y_i = -\cos(\pi i/M)$, $i = 0,...,M$, which include the interval ends. The collocation nodes for the interpolating expansions in the scaled Hermite functions are $Lx_0,...,Lx_N$, where $x_j$ are the roots of $H_{N+1}(x)$. These roots are spanned in the interval $[-\sqrt{2N}, \sqrt{2N}]$; to compute them we used the Matlab function *hermpts* from the free open-source package Chebfun [16]. Since

$$F[\Psi_j(x/L)] = L\sqrt{2\pi}(-i)^j \Psi_j(kL), \tag{19}$$

for collocation in the Fourier space it is convenient to use the grid of wave numbers $k_j = x_j/L$.

As the main variables in our spatially discretized problem, we choose the grid values $g_{i,j}(t) = g(Lx_j, y_i, t)$. To find the time derivatives $\dot{g}_{i,j}(t)$, we seek first an approximate solution $\tilde{g}$ to (13) in the form of an accounting for zero boundary conditions expansion in the Chebyshev polynomials of the second kind

$$\tilde{g}(k, y, t) = \sqrt{1-y^2} \sum_{m=0}^{M} \alpha_m(t,k) U_m(y) \tag{20}$$

with unknown expansion coefficients $\alpha_m(t,k)$. Since, see [10],

$$\frac{d}{dy}\left[U_m(y)\sqrt{1-y^2}\right] = -\frac{(m+1)T_{m+1}(y)}{\sqrt{1-y^2}} \tag{21}$$

we obtain

$$\partial_y \tilde{g} = -\sum_{m=0}^{M} \alpha_m(t,k)(m+1)\frac{T_{m+1}(y)}{\sqrt{1-y^2}}. \tag{22}$$

Let us consider the singular terms of equation (13) first. Using (22) and the formula (see [10])

$$\int_{-1}^{1} \frac{T_{m+1}(y')}{(y-y')\sqrt{1-y'^2}} dy' = -\pi U_m(y), \tag{23}$$

we find

$$\frac{1}{2\pi} \int_{-1}^{1} \frac{1}{y-y'} \partial_{y'} \tilde{g}(k, y', t) dy' = \frac{1}{2} \sum_{m=0}^{M} (m+1)\alpha_m(t,k) U_m(y). \tag{24}$$

To deal with the logarithmic kernel we use the known expansion [17], which converges in the weighted $L^2$-norm,

$$\ln|y-y'| = -\sum_{m=0}^{\infty} \tau_m T_m(y) T_m(y'), \tag{25}$$

where $\tau_0 = \ln 2$ and $\tau_m = 2/m$ for $m \geq 1$. Substituting, see [10], $T_0 = U_0, T_1 = U_1/2$, and $T_m = (U_m - U_{m-2})/2$ for $m \geq 2$, and assembling, we approximate $\ln|y-y'|$ by a bivariate expansion in the Chebyshev polynomials of the second kind with a sparse coefficient matrix,

$$\ln|y-y'| \approx \sum_{m,l=0}^{M} \lambda_{m,l} U_m(y) U_l(y').$$

Using the orthogonality relation (16) we obtain

$$\frac{1}{2\pi} \int_{-1}^{1} \ln\left(\frac{|k|}{2}|y-y'|\right) \tilde{g}(k, y', t) dy' = \frac{1}{4}\left\{\sum_{l=0}^{M}\left[\sum_{m=0}^{M} \lambda_{m,l}\alpha_l(t,k)\right] U_l(y) + \ln\left(\frac{|k|}{2}\right)\alpha_0(t,k)\right\}. \tag{26}$$

For each grid wave number $k = x_j / L$ the regular kernel parts in (13) can be approximated by the interpolating bivariate Chebyshev expansions (Appendix A):

$$\mathcal{A}(k, y - y') \approx \sum_{i,l=0}^{M} v_{i,l}(k) U_l(y') U_i(y),$$

$$\mathcal{B}(k, y - y') \approx \sum_{i,l=0}^{M} \gamma_{i,l}(k) T_l(y') U_i(y). \tag{27}$$

Using (15), (16), (20), (22), and (27), we find:

$$\int_{-1}^{1} \mathcal{A}(k, y - y') \tilde{g}(k, y', t) dy' = \sum_{m=0}^{M} \left[ \sum_{l=0}^{M} \eta_{m,l}(k) \alpha_l(t, k) \right] U_m(y), \tag{28}$$

$$\int_{-1}^{1} \mathcal{B}(k, y - y') \partial_x \tilde{g}(k, y', t) dy' = -\sum_{m=0}^{M} \left[ \sum_{l=0}^{M} \beta_{m,l} \alpha_l(t, k) \right] U_m(y), \tag{29}$$

where $\eta_{i,j} = \frac{\pi}{2} v_{i,j}$ and $\beta_{i,j} = \frac{\pi}{2}(j+1) \gamma_{i,j+1}$ with $\beta_{i,M} = 0$. Assembling (24), (26), (28), and (29), we arrive at the following equation for each grid wave number $k$,

$$\sum_{m=0}^{M} \vartheta_m(t, k) U_m(y) = -\tilde{Z}(k, y, t) \tag{30}$$

with

$$\vartheta_m = \sum_{l=0}^{M} \left( k^2 \eta_{m,l}(k) - \beta_{m,l}(k) - \frac{k^2}{4} \lambda_{m,l} \right) \alpha_l(t, k) +$$

$$\frac{(m+1)}{2} \alpha_m(t, k) - \delta_{0,m} \left( \frac{k^2}{4} \ln \frac{|k|}{2} \right) \alpha_0(t, k).$$

Denoting $\vec{\vartheta} = (\vartheta_0, ..., \vartheta_M)^T$, $\vec{\alpha} = (\alpha_0, ..., \alpha_M)^T$ we can write $\vec{\vartheta}(t, k) = S(k) \vec{\alpha}(t, k)$ with an $(M+1) \times (M+1)$ matrix $S(k)$.

We now compute the interpolating expansion in Chebyshev polynomials of the second type also for the right-hand-side of equation (30). First, using the known values $g_{ij}(t)$ we find an approximation to the grid values of $j_x = \partial_y g$ and $j_y = -\partial_x g$. This determines the grid values of $e = \rho(|j|) j$. The components $e_x, e_y$ should also be differentiated to find the grid values of $\nabla \times e = \partial_x e_y - \partial_y e_x$. In all cases, to differentiate numerically we use one of the two differentiation matrices, the $(M+1) \times (M+1)$ matrix $D^{Ch}$ or the $(N+1) \times (N+1)$ matrix $D^H$, obtained by differentiating, respectively, the Chebyshev polynomial and the Hermite function interpolating expansions (Appendix B), and such that

$$\left( \partial_y f(Lx_i, y_0), ..., \partial_y f(Lx_i, y_M) \right)^T = D^{Ch} \cdot \left( f(Lx_i, y_0), ..., f(Lx_i, y_M) \right)^T, \quad i = 0, ..., N,$$

$$\left( \partial_x f(Lx_0, y_j), ..., \partial_x f(Lx_N, y_j) \right)^T = D^H \cdot \left( f(Lx_0, y_j), ..., f(Lx_N, y_j) \right)^T, \quad j = 0, ..., M.$$

These matrices relate the grid values of the derivatives to the grid values of a function itself. Adding the grid values of $\dot{h}_z^e$ we obtain $Z = \dot{h}_z^e - \partial_x e_y + \partial_y e_x$ at the points $(Lx_i, y_j)$. We now construct the interpolating expansion $Z = \sum_{m,l=0}^{M,N} \kappa_{m,l} U_m(y) \Psi_l(x/L)$ (see Appendix A) and use (19) to find $\tilde{Z} = \sum_{m=0}^{M} \zeta_m(t,k) U_m(y)$ with $\zeta_m(t,k) = L\sqrt{2\pi} \sum_{l=0}^{N} (-\mathrm{i})^l \kappa_{m,l}(t) \Psi_l(Lk)$ for the grid wave numbers.

Equating expansions on the two sides of equation (30) we find $\vec{\alpha}(t,k_i) = -S^{-1}(k_i)\vec{\zeta}(t,k_i)$ and can calculate $\tilde{\dot{g}} = \sqrt{1-y^2}\sum_{m=0}^{M} \alpha_m(t,k) U_m(y)$ at the nodes $(k_i, y_j)$. Finally, using the interpolating expansions in the Hermite functions, we compute the inverse Fourier transform $\dot{g} = F^{-1}\left[\tilde{\dot{g}}\right]$ at the grid nodes, i.e., the time derivatives $\mathrm{d}g_{i,j}/\mathrm{d}t = \dot{g}(Lx_i, y_j, t)$, which are transferred to the ODE solver.

*3.2. Magnetic field computation.*

To account for a field-dependent current-voltage relation, e.g., a power-law relation with the sheet critical current density $j_c = j_c(\mathbf{h})$, it is needed to compute, for a known function $g(\mathbf{r},t)$, the grid values of the magnetic field $\mathbf{h} = \mathbf{h}^\mathrm{e} + \mathbf{h}^\mathrm{i}$, where $\mathbf{h}^\mathrm{i}$ is the field induced by the current $\mathbf{j} = \overline{\nabla}\times g$. The parallel to thin film component of the magnetic field is discontinuous; this component will be assumed equal to that of the applied field: $\mathbf{h}_\| = \mathbf{h}_\|^\mathrm{e}$. To find the normal component of the induced field, $h_z^\mathrm{i}$, we use the Hermite-Chebyshev expansions as follows.

First, using the matrices $D^{Ch}$, $D^H$, and the grid values of the stream function $g$, we find the grid values of current density components, then construct the weighted bivariate interpolating expansions

$$j_x(x,y,t)\sqrt{1-y^2} = \sum_{m,l=0}^{M,N} \varpi_{m,l}^x(t) T_m(y) \Psi_l(x/L),$$

$$j_y(x,y,t)\sqrt{1-y^2} = \sum_{m,l=0}^{M,N} \varpi_{m,l}^y(t) T_m(y) \Psi_l(x/L),$$

and use them to present the Fourier transform of these components as

$$\tilde{j}_x(k,y,t) = \frac{1}{\sqrt{1-y^2}} \sum_{m=0}^{M} \upsilon_m^x(t,k) T_m(y),$$

$$\tilde{j}_y(k,y,t) = \frac{1}{\sqrt{1-y^2}} \sum_{m=0}^{M} \upsilon_m^y(t,k) T_m(y)$$

with $\upsilon_m^x(t,k) = L\sqrt{2\pi} \sum_{l=0}^{N} \varpi_{m,l}^x(t)(-\mathrm{i})^l \Psi_l(kL)$ and $\upsilon_m^y(t,k) = L\sqrt{2\pi} \sum_{l=0}^{N} \varpi_{m,l}^y(t)(-\mathrm{i})^l \Psi_l(kL)$.

Applying the Fourier transform with respect to $x$ to the equation $h_z^\mathrm{i} = \nabla\times\int_\Omega G(\mathbf{r}-\mathbf{r}')\mathbf{j}(\mathbf{r}',t)\mathrm{d}\mathbf{r}'$ and separating again the singular kernel parts, we obtain an expression similar to that in (13) but now in terms of the current density components,

$$\tilde{h}_z^i = \int_{-1}^{1}\left\{\mathcal{B}(k,y-y') + \frac{1}{2\pi}\frac{1}{y-y'}\right\}\tilde{j}_x(k,y',t)\,dy' +$$

$$ik\int_{-1}^{1}\left\{\mathcal{A}(k,y-y') - \frac{1}{2\pi}\ln\left(\frac{|k|}{2}|y-y'|\right)\right\}\tilde{j}_y(k,y',t)\,dy'. \tag{31}$$

By (23) we find

$$\frac{1}{2\pi}\int_{-1}^{1}\frac{\tilde{j}_x(k,y',t)}{y-y'}dy' = -\frac{1}{2}\sum_{m=0}^{M}\upsilon_m^x(t,k)U_{m-1}(y),$$

where $U_{-1}=0$. Using (25) and the orthogonality relations (15) we obtain

$$\frac{1}{2\pi}\int_{-1}^{1}\tilde{j}_y(k,y',t)\ln\left(\frac{|k|}{2}|y-y'|\right)dx' =$$

$$-\sum_{m=0}^{M}\theta_m\upsilon_m^y(t,k)T_m(y),$$

where $\theta_m = \tau_m/4$ for $m>0$ and $\theta_0 = [\tau_0 - \ln(|k|/2)]/2$.

For each grid wave number $k = x_i/L$ here we approximate $\mathcal{A}$ and $\mathcal{B}$ by the interpolating expansions

$$\mathcal{A}(k,y-y') \approx \sum_{l=0}^{M}\hat{v}_l^{\mathcal{A}}(k,y)T_l(y'),$$

$$\mathcal{B}(k,y-y') \approx \sum_{l=0}^{M}\hat{v}_l^{\mathcal{B}}(k,y)T_l(y'), \tag{32}$$

and, for the regular integrals in (31), obtain

$$\int_{-1}^{1}\mathcal{A}(k,y-y')\tilde{j}_y(k,y',t)\,dy' = \sum_{m=0}^{M}\upsilon_m^y(t,k)v_m^{\mathcal{A}}(k,y)$$

$$\int_{-1}^{1}\mathcal{B}(k,y-y')\tilde{j}_x(k,y',t)\,dy' = \sum_{m=0}^{M}\upsilon_m^x(t,k)v_m^{\mathcal{B}}(k,y)$$

with $v_m^{\mathcal{A}} = \frac{\pi}{2}\hat{v}_m^{\mathcal{A}}$, $v_m^{\mathcal{B}} = \frac{\pi}{2}\hat{v}_m^{\mathcal{B}}$ for $m>0$ and $v_0^{\mathcal{A}} = \pi\hat{v}_0^{\mathcal{A}}$, $v_0^{\mathcal{B}} = \pi\hat{v}_0^{\mathcal{B}}$.

Assembling all terms in (31) we can find the grid values of

$$\tilde{h}_z^i = \sum_{m=0}^{M}\left[w_m^y(k,y)\upsilon_m^y(t,k) + w_m^x(k,y)\upsilon_m^x(t,k)\right],$$

where the grid values of $w_m^y(k,y) = ik\left[v_m^{\mathcal{A}}(k,y) + \theta_m T_m(y)\right]$, $w_m^x(k,y) = v_m^{\mathcal{B}}(k,y) - 0.5 U_{m-1}(y)$ are used on each time step but need to be calculated only once. Finally, using again the interpolating expansion in the Hermite functions, we compute the grid values of the induced magnetic field $h_z^i = F^{-1}[\tilde{h}_z^i]$.

As an example, we computed the field $h_z^i$ induced by a circular counter-clockwise current in the ring $r_1 \leq r \leq r_2$, where $r = |\mathbf{r}|$. We assumed $|\mathbf{j}| = (r_2 - r)(r - r_1)$ inside the ring and is zero outside. The stream function, satisfying equation $\mathbf{j} = \overline{\nabla} \times \mathbf{g}$, can be chosen as

$$g = \begin{cases} 0 & r \geq r_2, \\ (r_2 - r)^2(r_2 + 2r - 3r_1)/6 & r_1 < r < r_2, \\ (r_2 - r_1)^3/6 & r \leq r_1. \end{cases}$$

For $r_1 = 0.25$, $r_2 = 0.75$ the values of this function were assigned to nodes of the $61 \times 61$ Hermite-Chebyshev mesh with $L = 0.1$ (with this scaling the grid points $Lx_i$ are distributed in the interval $[-1.02, 1.02]$). The computed normal magnetic field component (Figure 1, left) is close to the axisymmetric 1D problem solution (Figure 1, right) obtained by numerical integration of the well-known formula [18] expressing the field of a circular current loop via elliptic integrals.

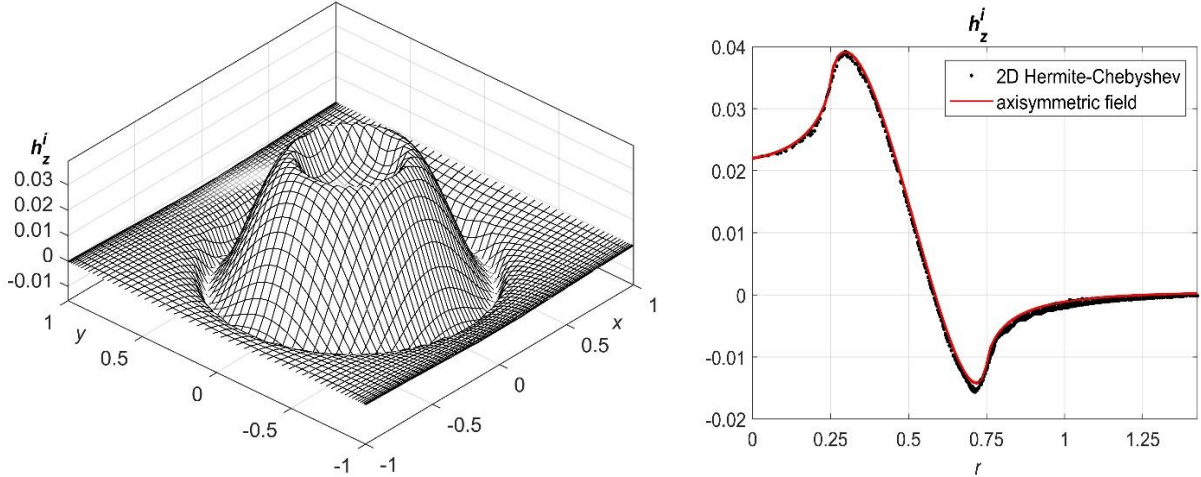

**Figure 1.** Left: magnetic field $h_z^i$ induced by the ring current, numerical simulation. Right: grid values of the 2D problem solution $h_z^i$ (black dots) plotted versus the node distance from the origin for a comparison with the semi-analytical axisymmetric field (red line). Scaled (dimensionless) variables.

### 4. Numerical simulation

We present two examples of the 2D problems solved by the Hermite-Chebyshev method: simulating an inhomogeneous strip with a transport current and modeling a superconducting dynamo flux pump.

*4.1. Inhomogeneous strip with transport current.*

Let the transport current be harmonic, $I = 200\sin(2\pi t/T)$ A, with $T = 0.02$ s (the frequency is 50 Hz). We assume the strip width $2a = 12$ mm, $\mathbf{h}^e = \mathbf{0}$, and $n = 20$ in the power current-voltage relation with the critical sheet current density $j_c = j_c^0 \chi^r(\mathbf{r}) \chi^h(h_z)$, where $j_c^0 = 23.6$ A/mm. Setting $\chi^r = 1 + \varphi(x+a, y+a) - \varphi(x-a, y-a)$ with $\varphi(x, y) = 0.75\exp\left(-[2x^2 + y^2]/a^2\right)$, we specify the spatial inhomogeneity of the strip in our example (indicated by the background colour in Figure 2, top). The function $\chi^h = 1/(1+h_0^{-1}|h_z|)$ describes the dependence on the magnetic field; in this example we assume $h_0 = 20 j_c^0$.

We solve the problem by simultaneously integrating in time the ODE system for the spatially discretized 1D problem (Appendix C), to find the far-away distributions $\boldsymbol{j}^\infty(y,t)$ and $\boldsymbol{e}^\infty(y,t)$, and the ODE system for the spatially discretized 2D problem (14) with $Z^* = \nabla \times (\boldsymbol{e} - \boldsymbol{e}^\infty)$, to compute the sheet current density $\boldsymbol{j} = \boldsymbol{j}^\infty + \overline{\nabla} \times g^*$ in the inhomogeneous strip (Figure 2).

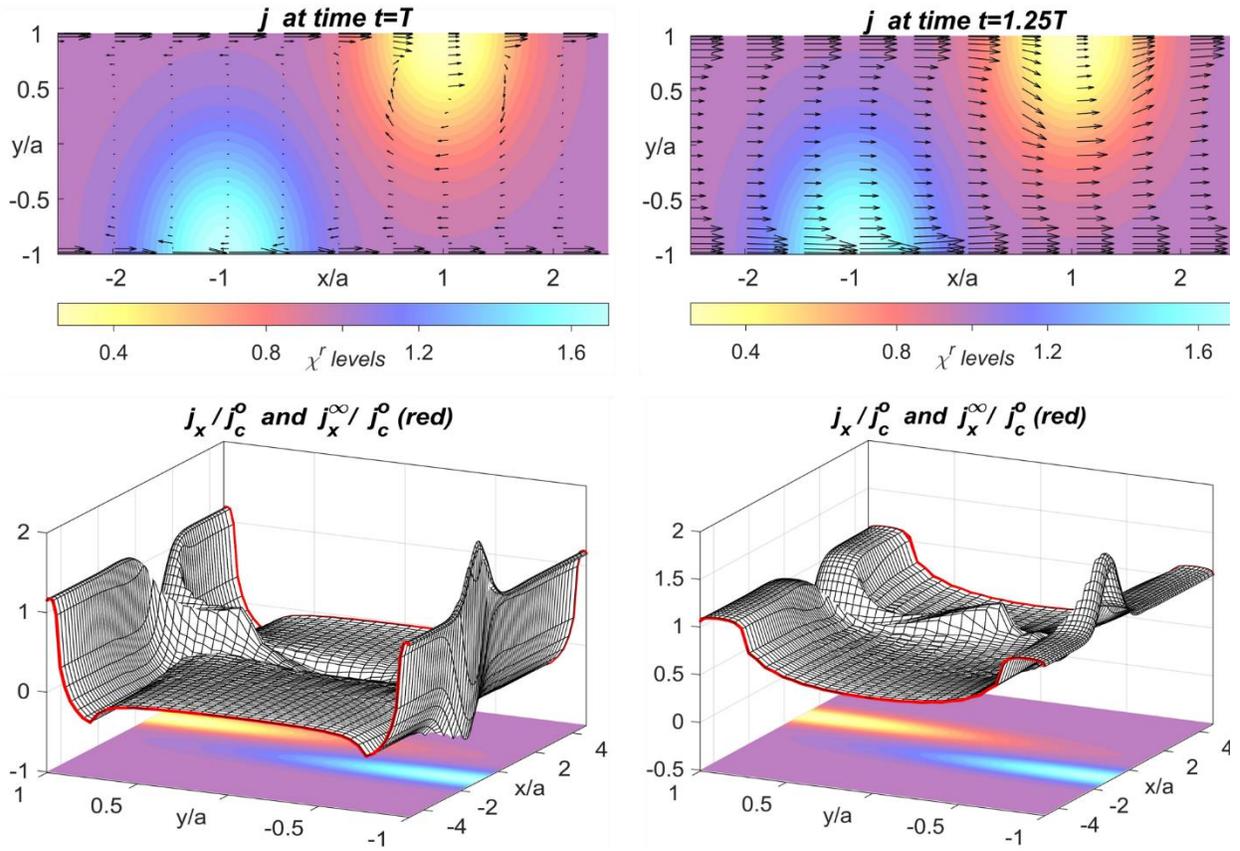

**Figure 2**. Inhomogeneous strip with the transport current, simulation results for $t = T$ ($I = 0$, left) and for $t = 1.25T$ ($I = 200$ A, right). Top: the current density distributions; the background color indicates spatial inhomogeneity of the critical sheet current density. Bottom: $j_x / j_c^0$ (surfaces) and $j_x^\infty / j_c^0$ (red lines).

In this example we use the Hermite-Chebyshev mesh with $M = 50$, $N = 80$ and the scaling factor $L = 0.4$. The Hermite nodes are spanned in the interval $-4.76a \leq x \leq 4.76a$. The influence of the strip inhomogeneity is negligible in strip points with the $x$-coordinate outside this interval: agreement between the 1D problem solution $j_x^\infty$ and the 2D problem solution $j_x$ is clearly observed near the interval ends (Figure 2, bottom). The distribution of the time-averaged loss power density, $p(\boldsymbol{r},t) = \boldsymbol{j} \cdot \boldsymbol{e}$, computed for the second cycle, is shown in Figure 3. As could be expected, the loss is much higher at the local depression of $j_c$ than in the other strip parts.

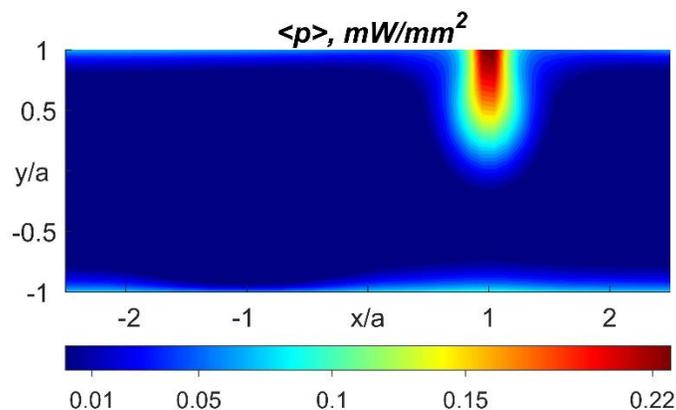

**Figure 3.** Time-averaged loss power density.

To study convergence of our numerical scheme, the computations were performed also using a cruder and a finer meshes. The scaling factors were such that the Hermite nodes of a finer mesh were spanned in a slightly larger interval than for a cruder one. Interpolating the finest mesh solution $j(x,y,t)$ onto two other meshes, we estimated the relative differences, $\delta j$, in the $L^1(\Omega \times [0,2T])$ norm (Table 1). Although computation time quickly increases for finer meshes, these estimates suggest that a good accuracy is achieved in a very reasonable time for the medium mesh ($M = 50, N = 80$). We note that, similarly to the known finite element methods for thin film problems, the sheet current density is not computed directly but is obtained by numerical differentiation of the auxiliary variable, the stream function, and this affects the accuracy.

**Table 1.** Numerical simulation results.

| Mesh | | | Computation | $\delta j$ |
|---|---|---|---|---|
| M | N | L | time per cycle | (%) |
| 25 | 40 | 0.5 | 36 sec. | 5.1 |
| 50 | 80 | 0.4 | 15 min. | 1.6 |
| 100 | 160 | 0.3 | 14 hours | --- |

*4.1 A high-temperature superconducting dynamo*

We now apply the Hermite-Chebyshev method to the 2D version [7] of the superconducting dynamo benchmark problem [6]. The external magnetic field $h^e$, acting upon the superconducting strip (dynamo stator), is induced by a uniformly magnetized permanent magnet (PM) having a rectangular prism shape and attached to a rotating dynamo rotor, see Figure 4. As in [7], this field is calculated as a rotational transformation of the analytical solution [19]. All problem parameters (Table 2) are as in [6, 7].

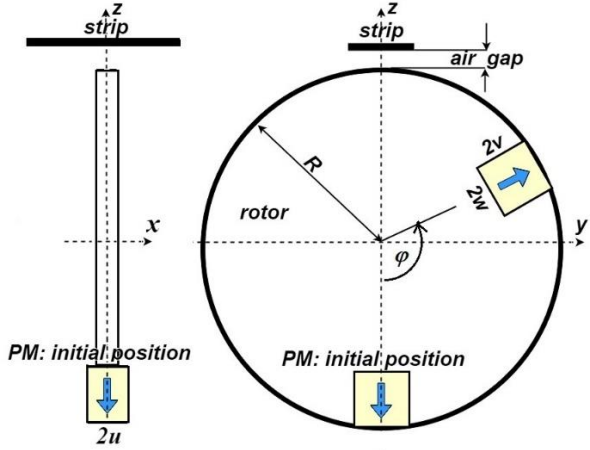

**Figure 4.** The superconducting dynamo, schematic drawing.

**Table 2.** Dynamo benchmark problem parameters

| | | |
|---|---|---|
| permanent magnet (PM) | width, $2w$ | 6 mm |
| | height, $2v$ | 12 mm |
| | length, $2u$ | 12.7 mm |
| | remanent flux density, $B_r$ | 1.25 T |
| stator strip | width, $2a$ | 12 mm |
| | critical sheet current density, $j_c^0$ | 23.6 A/mm |
| | power value, $n$ | 20 |
| rotor external radius, $R$ | | 35 mm |
| air gap, $d$ | | 3.7 mm |
| frequency of rotation, $f$ | | 4.25 Hz |

To compare the numerical simulation results obtained for this superconducting dynamo using the Hermite-Chebyshev method and those of the mixed finite element method [7], we computed by each method the width-averaged voltage $V(t)$ for two rotor rotations, $0 \leq t \leq 2/f$. As in [7], the power current-voltage relation with $j_c \equiv j_c^0$ and the open-circuit condition (zero transport current) were assumed.

Above, we have already used the interpolating bivariate expansion $e_x = \sum_{m=0}^{M} \sum_{l=0}^{N} e_{m,l}(t) U_m(y) \Psi_l(x/L)$. Substituting this into equation (9) we obtain

$$V(t) = \frac{1}{2} \sum_{m,l=0}^{M,N} e_{m,l}(t) \int_{-1}^{1} U_m(y) \mathrm{d}y \int_{-\infty}^{\infty} \Psi_l(x/L) \mathrm{d}x.$$

Here the Chebyshev polynomial integrals are [10]

$$\int_{-1}^{1} U_m(x) \mathrm{d}x = \begin{cases} \dfrac{2}{m+1} & m \text{ is even} \\ 0 & m \text{ is odd} \end{cases}$$

and for the Hermite functions are computed using the recurrence formula [14]:

$$\int_{-\infty}^{\infty} \Psi_0(x/L) \mathrm{d}x = L\pi^{1/4}\sqrt{2}, \quad \int_{-\infty}^{\infty} \Psi_1(x/L) \mathrm{d}x = 0,$$

$$\int_{-\infty}^{\infty} \Psi_{l+1}(x/L) \mathrm{d}x = \sqrt{\frac{l}{l+1}} \int_{-\infty}^{\infty} \Psi_{l-1}(x/L) \mathrm{d}x.$$

The mixed method was used with the finite element meshes in the strip domain $\Omega_0 = \{(x,y): |x| \leq 4a, \ |y| \leq a\}$, out of which the electric field was insignificant. To increase the accuracy, the meshes were refined in the central part of this domain. Furthermore, the second order finite difference approximation in time was used in this method with a varying time step, corresponding to the rotor rotation for 0.2° for rotor positions in which the permanent magnet is close to the strip and a much larger step otherwise. We present the data on convergence and computation times for each method in

Table 3, where the estimated errors $\delta V$ are the relative deviations (in the $L^1$ norm) from the $V(t)$ values calculated by the same method using the finest mesh.

**Table 3.** Comparison of two numerical methods.

| Mixed finite element method | | | Hermite-Chebyshev method | | | | |
|---|---|---|---|---|---|---|---|
| Mesh, number of elements | Time per cycle (min) | $\delta V$, % | Mesh | | | Time per cycle (min) | $\delta V$, % |
| | | | M | N | L | | |
| 1056 | 3.4 | 4.5 | 60 | 30 | 0.7 | 3.3 | 2.4 |
| 2180 | 33 | 1.6 | 90 | 46 | 0.6 | 31 | 0.9 |
| 4226 | 118 | --- | 120 | 60 | 0.5 | 186 | --- |

We note that the averaged voltage is an integral of the electric field which can be difficult to calculate accurately: due to the highly nonlinear current-voltage relation, small current density errors can lead to large errors in the electric field. The mixed method used in [7] was derived in [20, 21] for general thin superconducting film problems especially to resolve this difficulty. Nevertheless, in the considered thin strip problem, the Hermite-Chebyshev method turns out to be at least as accurate and efficient as the mixed method and, probably, outperforms it. The solutions obtained by these methods are close (Figure 5); this confirms that the chosen for the mixed method strip domain $\Omega_0$ is sufficiently long.

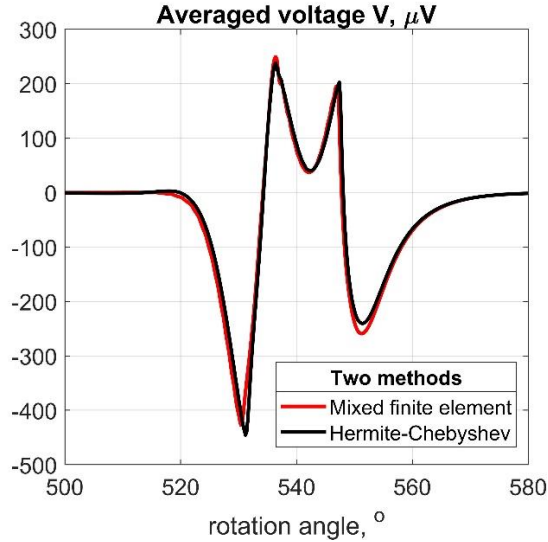

**Figure 5.** The averaged voltage during the second rotor rotation computed by two numerical methods.

We performed the open-circuit simulations also with the field-dependent critical sheet current density

$$j_c = \frac{j_c^0}{1 + h_0^{-1}\sqrt{h_z^2 + \kappa\left(h_x^2 + h_y^2\right)}}$$

with $\kappa = 0.5$ and $h_0 / j_c^0 = 5, 10,$ and $20$ (Figure 6); the field independent case corresponds to $h_0 / j_c^0 = \infty$. In these simulations we used the mesh with $M = 90, N = 46$ and computations took about twice more time because the magnetic field has had to be found on each time step. The current density distribution computed with $h_0 / j_c^0 = 10$ is shown for three rotor positions in Figure 7.

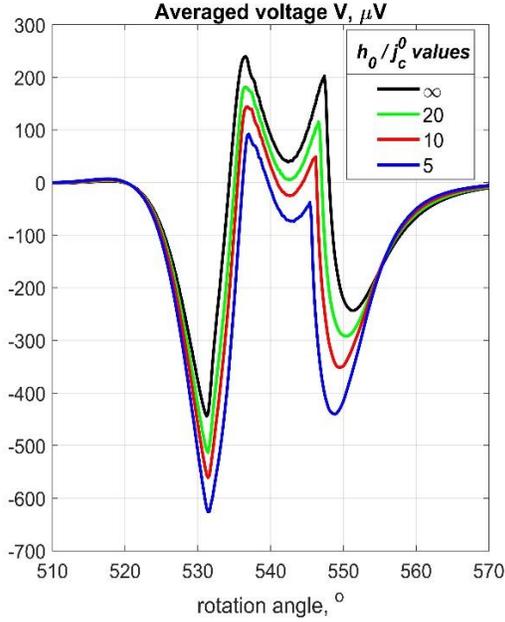

**Figure 6**. Averaged voltage, simulation results for field-dependent critical sheet current density.

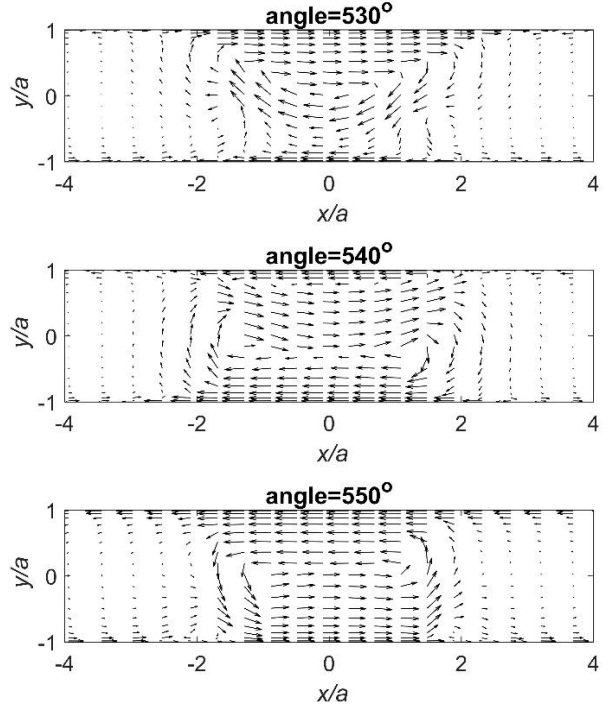

**Figure 7**. Sheet current density distribution for three rotor positions; $h_0 / j_c^0 = 10$.

## 5. Conclusion

Increasing production of coated conductors and their use in the HTS magnets, electrical machines, fault current limiters, etc., makes numerical solution of 2D superconducting strip problems an actual task. The derived pseudospectral Hermite-Chebyshev method can be applied to strip transport current and magnetization problems with an arbitrary current-voltage relation characterizing the superconducting material, to problems with spatially inhomogeneous strips and strips in a nonuniform applied field. Although application of spectral methods to superconductivity problems has started only recently [4, 5], in some cases these methods were shown to be more accurate and efficient than the finite element methods traditionally employed in this field. For a realistic 2D simulation of a superconducting dynamo, our method is fully competitive and, probably, more efficient than the advanced mixed finite element method [7] (see Table 3). Development of spectral methods for superconductivity problems seems a perspective direction of future research.

**Appendix A: Interpolating expansions in Chebyshev polynomials and Hermite functions.**

**A1)** Let a function $f(y)$ be interpolated at the points $y_0, ..., y_M$ by the expansions $\sum_{i=0}^{M} \alpha_i T_i(y)$, $\sum_{i=0}^{M} \beta_i U_i(y)$. Using the recurrence relations for the Chebyshev polynomials of the first and second type, see [10], one can efficiently construct the $(M+1) \times (M+1)$ matrices $P$ and $R$ with the elements, respectively, $P_{ij} = T_j(y_i)$, $R_{ij} = U_j(y_i)$, $i,j = 0,...,M$, so that $\vec{f} = P\vec{\alpha}$, $\vec{f} = R\vec{\beta}$, where $\vec{f} = (f(y_0),...,f(y_M))^T$. The inverse matrices $P^{-1}$ and $R^{-1}$ can be used to find the vectors of expansion coefficients, $\vec{\alpha}$ and $\vec{\beta}$ for a known vector $\vec{f}$.

Similarly, let the expansion $\sum_{i=0}^{N} \gamma_i \Psi_i(x/L)$ interpolate $f(x)$ at the points $Lx_i$. Then $\vec{f} = Q\vec{\gamma}$ and $\vec{\gamma} = Q^{-1}\vec{f}$, where $\vec{f} = (f(Lx_0),...,f(Lx_N))^T$ and the $(N+1)\times(N+1)$ matrix $Q$ has elements $Q_{ij} = \Psi_j(x_i)$, efficiently calculated using the recurrence relations (17). Note that with the wave number grid $k_j = x_j/L$ we can use the same matrices, $Q$ and $Q^{-1}$, for the interpolating expansions $\sum_{i=0}^{N} \gamma_i \Psi_i(kL)$ in the Fourier space.

**A2)** We can now construct also the bivariate Chebyshev and Hermite-Chebyshev interpolating expansions employed in this work. Let a function $F(y, y')$ be represented by the $(M+1)\times(M+1)$ matrix $\bar{F}$ with elements $\bar{F}_{i,j} = F(y_i, y_j)$. Then, e.g., the columns of $P^{-1}\bar{F}$ are coefficients of the interpolating expansion in polynomials $T_k(y)$ for the corresponding columns of $\bar{F}$. The bivariate interpolating expansions of $F$, $\sum_{i,j=0}^{M} \alpha_{ij} T_i(y)T_j(y')$, $\sum_{i,j=0}^{M} \beta_{ij} U_i(y)T_j(y')$, and $\sum_{i,j=0}^{M} \gamma_{ij} U_i(y)U_j(y')$, have, respectively, coefficient matrices $\bar{\alpha} = P^{-1}\bar{F}(P^{-1})^T$, $\bar{\beta} = R^{-1}\bar{F}(P^{-1})^T$, and $\bar{\gamma} = R^{-1}\bar{F}(R^{-1})^T$.

Similarly, let a function $F(x, y)$ be represented by the $(M+1)\times(N+1)$ matrix $\bar{F}$ with elements $\bar{F}_{i,j} = F(Lx_j, y_i)$. Then, the interpolating expansions $\sum_{i,j=0}^{M} \chi_{ij} T_j(y)\Psi_i(x/L)$ and $\sum_{i,j=0}^{M} \eta_{ij} U_j(y)\Psi_i(x/L)$ have, respectively, the coefficient matrices $\bar{\chi} = P^{-1}\bar{F}(Q^{-1})^T$ and $\bar{\eta} = R^{-1}\bar{F}(Q^{-1})^T$.

**Appendix B: Differentiation matrices $D^{Ch}$ and $D^H$.**

**B1)** Let $\sum_{k=0}^{M} \alpha_k T_k(y)$ interpolates $f(y)$ at $y_0,...,y_M$. Then $\vec{\alpha} = P^{-1}\vec{f}$ (Appendix A) and, since $dT_k(y)/dy = kU_{k-1}(y)$, we set $df/dy \approx \sum_{k=1}^{M} k\alpha_k U_{k-1}(y)$. Introducing the sparse $(M+1)\times(M+1)$ matrix $C$ with nonzero elements $C_{k,k+1} = k+1$, $k = 0,...,M-1$, we can write the coefficients of the latter expansion as $CP^{-1}\vec{f}$. The grid values of this approximation to $df/dy$ form (see Appendix A) the vector $\vec{f}' = RCP^{-1}\vec{f}$, so the Chebyshev differentiation matrix $D^{Ch} = RCP^{-1}$.

**B2)** To construct the Hermite differentiation matrix $D^H$ we start with the expansion $\sum_{k=0}^{N} \alpha_k \Psi_k(x/L)$ interpolating $f(x)$ at the grid points $Lx_0,...,Lx_N$. Now we have $\vec{\alpha} = Q^{-1}\vec{f}$ (Appendix A). Since, see [14],

$$d\Psi_k(u)/du = \sqrt{\frac{k}{2}}\Psi_{k-1}(u) - \sqrt{\frac{k+1}{2}}\Psi_{k+1}(u),$$

we set $df/dx \approx (1/L)\sum_{k=0}^{N} \beta_k \Psi_k(x/L)$ with $\beta_k = \left(\sqrt{k+1}\alpha_{k+1} - \sqrt{k}\alpha_{k-1}\right)/\sqrt{2}$, where $\alpha_{N+1}$ and $\alpha_{-1}$ should be set to zero. Hence, $\vec{\beta} = B\vec{\alpha}$, where $B$ is the sparse $(N+1)\times(N+1)$ matrix

$$B_{ij} = \frac{1}{\sqrt{2}}\begin{cases} -\sqrt{i} & j=i-1, \\ \sqrt{i+1} & j=i+1, \\ 0 & \text{otherwise.} \end{cases}$$

The grid values of $df/dx$ are, therefore, approximated as $\vec{f}' = (1/L)QB\vec{\alpha}$, so $D^H = (1/L)QBQ^{-1}$.

**Appendix C: Spectral solution of the 1D problem.**

In the 1D problem, $\boldsymbol{j}$ and $\boldsymbol{e}$ are directed along the $x$ axis and can be regarded scalar. Let a nonlinear current-voltage relation $e = \rho(j,\boldsymbol{h})j$, the transport current $I(t)$, and a uniform external field $\boldsymbol{h}^e(t)$ be given. In the dimensionless variables, the integrodifferential equation to be solved is, see [5],

$$\frac{1}{2\pi}\int_{-1}^{1}\frac{\partial_t j(y',t)dy'}{y-y'} = \partial_y e - \dot{h}_z^e. \tag{33}$$

To comply with our scheme for the 2D problem we need to use the same mesh $y_0, \ldots, y_M$, which includes the interval ends $y=\pm 1$. This does not allow us to solve this equation for $\partial_t j$ by the method [5], not applicable for this set of Chebyshev nodes. We overcome this difficulty by changing our main variable and using an analogue of the stream function.

Denoting $j_0 = j - I/2$, here we define $g(y,t) = \int_{-1}^{y} j_0(y,t)dy$. Then $j = \partial_y g + I/2$ and $g(\pm 1,t) = 0$. Let at time $t$ the values $g_k(t) = g(y_k,t)$ be known. Differentiating the interpolating Chebyshev expansion, we can calculate the node values of $j$ (Appendix B). If the nonlinear resistivity $\rho$ depends on the magnetic field, to find the electric field we need to find

$$h_z = h_z^e + \frac{1}{2\pi}\int_{-1}^{1}\frac{j(y',t)dy'}{y-y'}.$$

By constructing the interpolating expansion $j(y,t)\sqrt{1-y^2} = \sum_{l=0}^{M} v_l T_l(y)$ and using (23), we obtain

$$\frac{1}{2\pi}\int_{-1}^{1}\frac{j(y',t)dy'}{y-y'} = -\frac{1}{2}\sum_{l=1}^{M} v_l U_{l-1}(y).$$

The mesh values of $e$, and then also of $\partial_y e - \dot{h}_z^e$, the right-hand-side of (33), can now be found.

Let us seek $\dot{g}$ in the form $\dot{g} = \sqrt{1-y^2}\sum_{k=0}^{M}\alpha_k(t)U_k(y)$. Using (21) and (23), we obtain

$$\partial_t j(y) = \dot{I}/2 - \frac{1}{\sqrt{1-y^2}}\sum_{k=0}^{M}(k+1)\alpha_k(t)T_{k+1}(y),$$

$$\frac{1}{2\pi}\int_{-1}^{1}\frac{\partial_t j(y',t)dy'}{y-y'} = \frac{1}{2}\sum_{k=0}^{M}(k+1)\alpha_k(t)U_k(y) - \frac{\dot{I}}{4\pi}\bigl[\ln|y-1|-\ln|y+1|\bigr].$$

Making use of (25), we approximate

$$\ln|y-1|-\ln|y+1| \approx \phi^M(y) = -\sum_{k=0}^{M}\tau_k T_k(y)[T_k(1)-T_k(-1)]$$

with $T_k(1)-T_k(-1)=2$ if $k$ is odd and zero if $k$ is even, and arrive at

$$\frac{1}{2}\sum_{k=0}^{M}(k+1)\alpha_k(t)U_k(y) = \partial_y e - \dot{h}_z^e + \frac{\dot{I}}{4\pi}\phi^M(y).$$

Interpolating the mesh values of the right-hand-side of this equation by an interpolating expansion in $U_k$ and comparing the coefficients, we find $\alpha_k$. This determines $\dot{g}$ in the mesh nodes and the method of lines can be employed.